\documentclass[twocolumn,twoside,slac_two]{revtex4}
\usepackage{graphicx}
\usepackage{fancyhdr}
\usepackage{hyperref}
\hypersetup{
    colorlinks=true,
    urlcolor=magenta,
}
\begin{document}

\title{Correlation of AOT with Relative Frequency of Air Showers with energy 10$^{15}$ - 10$^{16}$ eV by Yakutsk Data}

%

\author{S. Knurenko, I. Petrov}
\affiliation{Yu. G. Shafer Institute of Cosmophysical Research and Aeronomy SB RAS, Yakutsk, Russia}

\begin{abstract}
Long-term series of measurement of spectral transparency of the atmosphere ($\lambda$ = 430 nm) and atmospheric optical thickness (AOT) measured by multimode photometer CE 318 in the region of Yakutsk array are analyzed ~\cite{PetrovIS-bibref1}. Correlation of AOT with intensity of air showers with small energies 10$^{15}$ - 10$^{16}$ eV is found.
The variability of aerosol composition of the atmosphere during the registration period of the Cherenkov light should be taken into account since it may affect the quality of determining characteristics of air showers ~\cite{PetrovIS-bibref2}.

\end{abstract}

\maketitle

\thispagestyle{fancy}


\section{Introduction}\label{PetrovIS-intro}

Polar regions of Yakutia, which can be attributed to the neighborhoods of Yakutsk (N 61$^\circ$39$^\prime$, E 129$^\circ$22$^\prime$, Alt 118 m), from the geophysics point of view, is interesting with anomalous manifestations in the upper atmosphere above 80 km above sea level: the invasion of flux of electrons ( auroras, radio noises etc.), their interactions with the environment, with the loss of significant energy and further the mechanism or mechanisms of energy transfer in the lower atmosphere. To some extent, these processes together with galactic cosmic rays may be involved in the formation of weather in the world ~\cite{PetrovIS-bibref1, PetrovIS-bibref2}. In ~\cite{PetrovIS-bibref3, PetrovIS-bibref4}, it was considered that the effect of solar and galactic cosmic rays on the temperature, therefore the weather and climate of our planet is practically proven.

Weather and climate in the region and on a global scale, of course, play the role of factors such as man-made disasters, volcanic activity, forest fires and the global greenhouse effect. To track the changes in atmospheric parameters stations network was created, placed on all continents. The main objective of this project was the global monitoring of the atmosphere and the identification of the causes of changes in parameters of the atmosphere, which could affect the Earth's climate as a whole. Yakutsk is one of the points in the global station circuit that supplies such information ~\cite{PetrovIS-bibref5}.

In this paper, one of the directions indicated by the data collection system independent of optical instruments that monitor the state of the atmosphere ~\cite{PetrovIS-bibref6} and presents data on the optical thickness of the atmosphere (AOT) in the context of long-term observations.

\section{The Yakutsk array}

\begin{figure}
\includegraphics[width=0.8\linewidth]{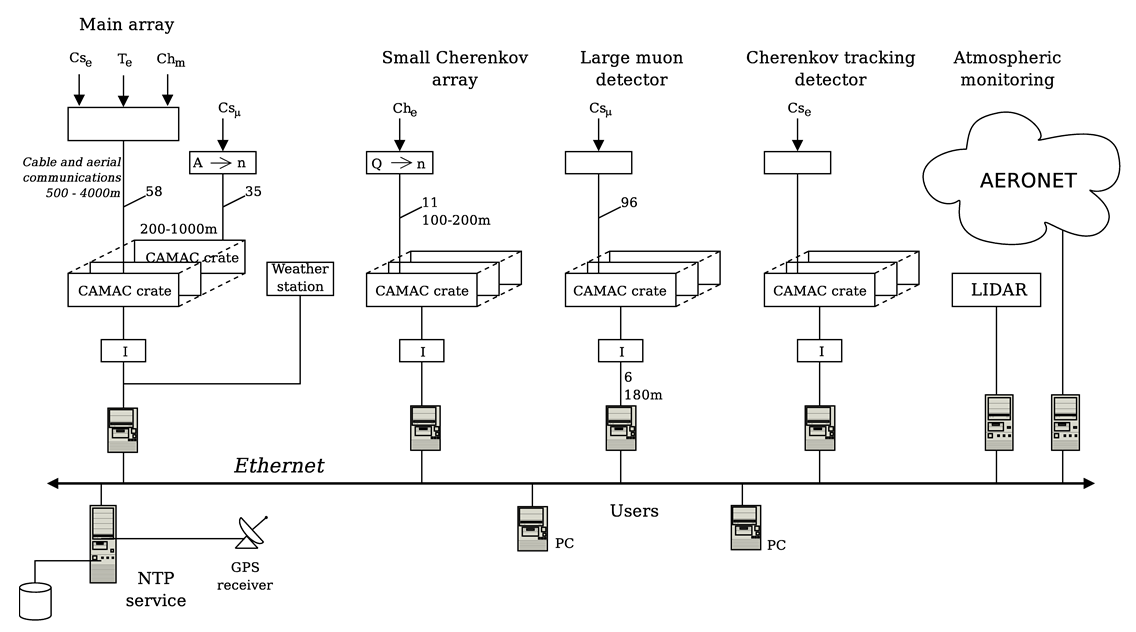}
\caption{Flowchart of the Yakutsk array LAN}
\label{PetrovIS-fig1}
\end{figure}

The Yakutsk array consists of a network of stations for registration using scintillation and Cherenkov detectors of different types of elementary particles: hadrons, electrons, positrons, muons and Cherenkov photons ~\cite{PetrovIS-bibref7}. The measurements were synchronized using GPS systems and form a single LAN of the Yakutsk array. The scheme of the network is shown in Fig.\ref{PetrovIS-fig1}.

Regular observations of the atmosphere at the Yakutsk started in 1970 and extended continuously to the present. With automatic mini stations measure temperature, pressure, and humidity. Particular attention was paid to the winter period of observation because at this time in the Yakutsk were observing the Cherenkov radiation in the optical wavelength range, and therefore required assessment of the state of the atmosphere. At the first stage, the transparency was determined by tracking the stars in the sky (in the area of the North Star), later on, the measurement repetition rate "Cherenkov flashes" in the atmosphere ~\cite{PetrovIS-bibref8}. Since 2004 he has been involved, and the laser at a wavelength of 532 nm, and a multimode photometer CE 318 for monitoring the atmosphere ~\cite{PetrovIS-bibref6, PetrovIS-bibref9}. Based on these measurements formed the base of the atmosphere state data.

\begin{figure}
\includegraphics[width=0.8\linewidth]{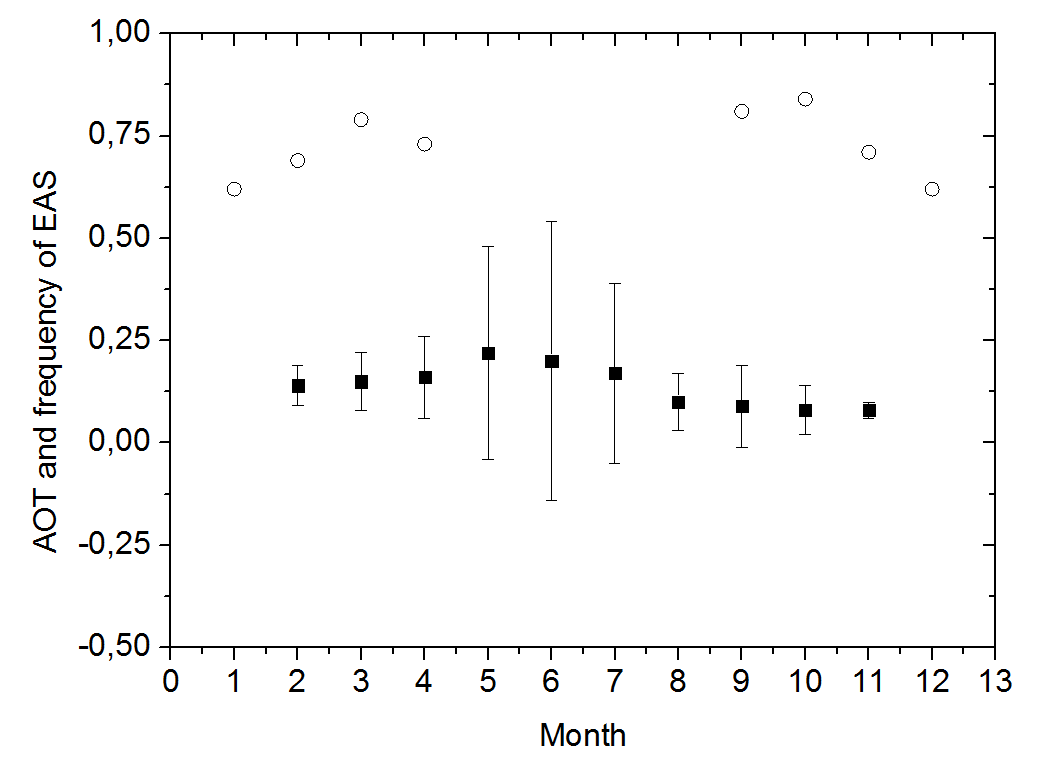}
\caption{AOT parameter comparison with relative frequency of air showers with energy 10$^{15}$-10$^{16}$ eV}
\label{PetrovIS-fig2}
\end{figure}

\section{Atmospheric Optical Thickness (AOT)}

Aerosol atmosphere - a slurry of solid and liquid microparticles are classified as the height and on the geographical area. This classification allows one to select specific sources of atmospheric aerosol and aerosol basic transformation processes under the influence of geophysical factors, including cosmic radiation and galactic origin ~\cite{PetrovIS-bibref10, PetrovIS-bibref2}.

It is believed that in Yakutia, the atmosphere is more clear than in regions with developed industrial structure ~\cite{PetrovIS-bibref11}. On top of Yakutia is located in the area with a sharply continental climate, when the air temperature in winter can reach -55 $^\circ$C, and in summer + 35 $^\circ$C. This drastic restructuring of the atmosphere in the winter months is accompanied by a temperature inversion, which leads to an increase in the density of the air surface layers of the atmosphere and the frosty fog ~\cite{PetrovIS-bibref6, PetrovIS-bibref16}.

In the summer months, there is an increase AOT due to smoke the atmosphere due to forest fires. Fig.\ref{PetrovIS-fig2} presents data on the AOT parameter, averaged monthly units for the observation period from 2002 and 2013 ~\cite{PetrovIS-bibref5}. Fig. \ref{PetrovIS-fig2} shows that even averaged over many years, and these tend to vary during the year that observed seasonal variation. In the summer months, AOT more than in the autumn and spring seasons and it is connected with the intensification of cyclonic activity and the influx of large masses of the region's water spray. On the other hand, in the summers of numerous forest fires are recorded almost every year, the smoky fraction of covering vast areas in southern and central parts of Yakutia, which significantly affects the composition of the aerosol. In autumn and spring, cyclical activity in the area of Yakutsk reduced - this is due to the stabilization processes in the atmosphere with the arrival of moderately low and freezing temperatures during these periods AOT has minimal value. In winter, the AOT measurement is rare, because at very low temperatures occur near ground frosty haze that interferes with measurements at the low standing sun. Since the intensity of the radiation in the optical range of EAS depends on the state of the atmosphere, it is interesting to find out whether there is a correlation between AOT parameter with frequency "Cherenkov flashes" small EAS. For this purpose, the measurement data of both the characteristics of the joint for several years were analyzed.

\section{The frequency of cosmic rays with an energy of 10$^{15}$ - 10$^{16}$ eV}

Relativistic particles passing through the atmosphere to form a radiation of different nature, including in the optical spectrum. Most of them are intense ionization glow of ionized nitrogen atoms and Cherenkov radiation, which are used mainly for the study of the longitudinal development of the EAS. As a first approximation, we can assume there is a powerful "flash" light, which recorded fast PMT within 10$^{-8}$ s. In a clean atmosphere, such phenomena are recorded without distortion. By registering the EAS Cherenkov light in different periods of observation and cool considering the incident energy spectrum of EAS (integral index is $\gamma$ = - 2), the frequency of such showers will characterize the state of the atmosphere. From a purely Rayleigh, with only molecular scattering, to turbid, with a high content of various types of aerosol. Tracking time rate "bursts" with respect to the night with the purest atmosphere, one can judge the transparency of the atmosphere ~\cite{PetrovIS-bibref8}.

To register EAS with energies 10$^{15}$ - 10$^{16}$ eV in Yakutsk used small Cherenkov array  ~\cite{PetrovIS-bibref7}, which selects the showers on the "Cherenkov trigger" - coincidence signals during 2.5 microseconds from three Cherenkov detector placed at the vertices of equilateral triangles with side of 50 m, 100 m and 250 m. Registration showers and management of the Cherenkov array by using the software package ~\cite{PetrovIS-bibref12}. A more detailed analysis of the technique described in ~\cite{PetrovIS-bibref13, PetrovIS-bibref14}.

Fig. \ref{PetrovIS-fig2} shows the joint observations between 2004 and 2013. We consider the periods of time during which the observations were made of the Cherenkov light EAS (September - April). From Fig. \ref{PetrovIS-fig2}, it follows that the EAS frequency increases with decreasing values of AOT in spring - autumn periods and a few falls in the winter months with the lowest temperatures from November to February ~\cite{PetrovIS-bibref15}. You may notice a correlation in the experimental data that suggests the possibility of a first approximation, doing assessment AOT frequency "Cherenkov flashes" from small EAS in those periods of time when the last measurement is not possible. Thus, using the method of measuring the frequency of the EAS, one can monitor the status of the atmospheric boundary layer. That is the conclusion we come in ~\cite{PetrovIS-bibref8, PetrovIS-bibref16}, which compares transparency, obtained by registering the EAS Cherenkov light and transparency, as measured with a laser at a wavelength $\lambda$ = 532 nm ~\cite{PetrovIS-bibref9}.

\section{Conclusion}

In carrying out optical measurements, EAS is important to control such features as the AOT and the absolute transparency of the atmosphere. It has been observed that in areas with an extreme continental climate, which include Yakutia, geophysical and climatic conditions contribute to the structure of the atmospheric model. Consequently, the change of parameters of the atmosphere also affects the quality of the registration of Cherenkov light, and in the future on energy estimates power EAS ~\cite{PetrovIS-bibref17}. Preliminary we can say:

\begin{enumerate}
\item On the monthly mean values of AOT in Yakutsk area may affect the restructuring of the atmosphere during the transition from autumn to winter and from winter to spring. This factor can influence the average monthly values of AOT in the measurement of this parameter in the winter time;
\item AOT and the number of "Cherenkov flashes" are correlated with each other and can be used as indicators of purity (transparency) of the atmosphere.

\end{enumerate}

\bigskip 

\end{document}